\documentclass[structabstract]{aa}  
\usepackage{graphicx}
\usepackage{txfonts}
\usepackage{threeparttable}
\usepackage{ulem}
\def\new#1 {{\bf #1} } \def\cut#1 {\sout{#1}}
\begin{document}
%
%   \title{Molecular observations of the starburst galaxy 
%          Henize~2-10}\thanks{Just to show the usage
%          of the elements in the author field}

   \title{Resolving the molecular environment of Super Star Clusters
       in Henize~2--10
      }

%   \subtitle{bla}

   \author{G. Santangelo
          \inst{1,2,3}
          \and
          L. Testi
          \inst{1,4}
          \and
          L. Gregorini
          \inst{2,3}
          \and
          S. Leurini
          \inst{1}
          \and
          L. Vanzi
          \inst{5}
          \and
          C.M. Walmsley
          \inst{4}
          \and
          D.J. Wilner
          \inst{6}
          }

   \institute{European Southern Observatory,
              Karl Schwarzschild str.2, D-85748 Garching bei Muenchen, Germany\\
              \email{gsantang@eso.org, ltesti@eso.org}
         \and
             INAF - Istituto di Radioastronomia, via Gobetti 101,
             40129 Bologna, Italy
         \and
             Dipartimento di Astronomia, Universit\`a di Bologna,
             via Ranzani 1, 40127 Bologna, Italy
         \and
             INAF - Osservatorio Astrofisico di Arcetri, Largo E. Fermi 5,
             I-50125 Firenze, Italy
         \and
             Pontificia Universidad Catolica de Chile,
             Departamento de Ingenieria Electrica,
             Av. Vicu~na Mackenna 4860,
             Santiago, Chile
         \and
             Harvard-Smithsonian Center for Astrophysics, 60 Garden Street,
             Cambridge, MA 02138, USA
             }

   \date{...}

% \abstract{}{}{}{}{} 
% 5 {} token are mandatory
 
  \abstract
  % context heading (optional)
  % {} leave it empty if necessary  
   {The rate of star formation both in the Galaxy and in external galaxies
should be related to the physical properties of the molecular clouds from which
stars form. This is expected  for the starbursts found both in
irregular galaxies and in some mergers.  The dwarf galaxy Henize~2-10 is
particularly interesting in this context as it shows a number of newly formed
Super Star Clusters (SSCs) associated with a very rich molecular environment.}
  % aims heading (mandatory)
{We present a high angular resolution study of the molecular gas associated
with the SSCs with the aim of deriving the physical properties of the 
parent molecular clouds. The final goal is to test the expectation that 
the formation of SSCs requires exceptionally dense and massive clouds.}
  % methods heading (mandatory)
   {We have used the  Submillimeter Array with an angular resolution
of $1.\!\!^{\prime\prime}9 \times 1.\!\!^{\prime\prime}3$ to map the J=2-1
transition of CO in Henize~2-10. Supplementary
measurements of HCN($J$=1-0), $^{13}$CO($J$=2-1) and millimeter
continuum were obtained with the APEX, IRAM~30m and SEST single dish telescopes.}
  % results heading (mandatory)
   {Our single dish observations confirm the association of the newly formed
SSCs in Henize~2-10 with dense molecular gas. Our interferometric observations
resolve the CO(2-1) emission in several giant molecular clouds. Overall 
the molecular gas accounts for approximately half of the mass in the central
regions of Henize~2-10. Although we find indications that the molecular clouds
associated with the formation of SSCs in Henize~2-10 are massive and dense, the
tracer we used (CO) and the linear resolution of our observations
(60$\times$80~pc) are still not adequate to test the expectation that
exceptionally dense and massive cores are required for SSCs formation.}
  % conclusions heading (optional), leave it empty if necessary 
   {}

   \keywords{Galaxies: dwarf -- Galaxies: individual: Henize~2-10 -- Galaxies: starburst
               }

%%%\titlerunning{Resolving the molecular environment of SSCs in Henize~2--10}
%%%\authorrunning{Santangelo et al.}
   \maketitle
%
%________________________________________________________________

\section{Introduction}

Galactic and extragalactic star formation mostly occurs in "bursts" or
clusters and the majority of stars is thought to form in large
clusters. However, cluster masses and densities vary greatly even within the
Milky Way (see Lada \& Lada 2003)
where the  well studied  Orion nebular cluster (ONC)
has been estimated to have a mass of order 1800 $M_{\odot}$ (Hillenbrand \& Hartmann 1998)
while the most massive
galactic clusters (e.g. Brandner et al. 2008; Dowell et al. 2008)
have masses of a few times $10^4 \, M_{\odot}$.
These are dwarfed however by young extragalactic clusters and in particular
by the starbursts or super star clusters (SSCs) seen in merging systems (e.g. Whitmore 2002)
and in some irregular galaxies (e.g. Elmegreen 2002; Johnson \& Kobulnicky 2003)
where masses can reach  $10^6 \,  M_{\odot}$. It has been speculated that
in this case one may be observing young globular clusters.

One presumes that the diversity between local and distant star
formation  reflects variation within the molecular
clouds from which these clusters form and, in particular, variations in
the gas pressure. Indeed, even locally, one observes large differences
in the characteristics of nearby clouds with moderate pressures and
densities in Taurus cores, which mostly form isolated stars,
and higher pressures in
the molecular gas in the Orion cloud neighbouring the ONC.  Moreover,
both gas pressures and star formation rates are  higher  in
the inner regions of the Milky Way, where most of the young massive star clusters
are located, and in external galaxies (e.g. Kennicutt 1998).
A summary of the characteristics of both
clouds and clusters is given by Tan~(2008).

Of particular interest are the youngest extragalactic embedded
super star clusters or SSCs which have been
identified by the  free-free radio emission from their associated
HII regions (e.g. Kobulnicky \& Johnson 1999; Tarchi et al. 2000;
Johnson \& Kobulnicky 2003).
These are similar in many ways to local ultra--compact HII regions
but larger and are typically powered by the equivalent of
several thousand O stars (e.g. Kobulnicky \& Johnson 1999). It is
estimated that their stellar masses  exceed $\sim \, 10^5 \, M_{\odot}$,
their radii are of order of a few parsecs, their electron densities are of order of 
$10^4$  cm$^{-3}$ and their ages are below 1 Myr (Kobulnicky \&
Johnson 1999; Vacca et al. 2002; Johnson \& \ Kobulnicky 2003).    Thus
their mass is of the same order as galactic GMCs but their dimensions are
smaller by an order of magnitude and, correspondingly their densities
and pressures are much larger. 
%Presumably also the star formation efficiency is less than unity and hence the molecular cloud
%progenitors of the SSCs are  more massive than the values given above.
For an assumed star forming efficiency of around 10\%, we might expect to see 
molecular forerunners of the SSCs with masses of $10^6 \, M_{\odot}$ and sizes of a few parsecs. 
One might hope to detect such precluster
``cores'' either in molecular line emission or in continuum
dust emission.

The dwarf irregular galaxy Henize 2-10 at a distance of
9 Mpc ($H_0$=75 km~s$^{-1}$~Mpc$^{-1}$,\cite{vacca92}; Kobulnicky \& Johnson 1999) 
is a good example of this phenomenon. It contains
several times $10^8 \, M_{\odot}$ of molecular gas as
well as multiple HII regions and young clusters observed at  radio, optical,
and infrared wavelengths (Kobulnicky \& Johnson 1999; Johnson
et al. 2000; Vacca et al. 2002; Cabanac et al. 2005).  The
nebulosity seen in HST images does not coincide well with the
radio knots suggesting large amounts of obscuration.  It is clear
therefore that higher angular resolution observations of the molecular
emission are warranted in order to understand the characteristics of the
clouds from which the SSCs form. Ultimately of course, one would
like to image  molecular line emission with a resolution comparable
both to the dimensions of the SSCs and the optical images (0.1
arcsec resolution).  This is currently not possible but
as  a first step in this direction, we present
here results obtained with the Submillimeter array (SMA)
in the CO J=2-1 transition  (and in the mm continuum)
with a resolution of roughly 1$.\!\!^{\prime\prime}$8
(80 parsec at the distance of Henize~2-10). This allows us to
see the distribution of structures  comparable to galactic GMCs and
to compare with images at other wavelengths. It also allows us
an improved view of the molecular cloud dynamics than available
in earlier work (Kobulnicky et al. 1995).  Finally, we have obtained 
supplementary measurements of the millimeter continuum and the
HCN($J$=1-0) and $^{13}$CO($J$=2-1) lines with the SEST, IRAM~30m and
APEX single dish telescopes.

This paper is structured as follows. In section \ref{observations}, 
we present our observations. In section \ref{results}, we present our results and
compare with previous CO observations of Henize~2-10. In
section \ref{derivation}, we describe our analysis and derive the physical 
parameters of the detected clouds. In section \ref{discussion}, we discuss the results and finally, 
in section \ref{summary}, we present our summary and conclusions.

%__________________________________________________________________

\section{Observations and Data Reduction}\label{observations}

\subsection{Single-dish observations}

\subsubsection{IRAM-30m Telescope}
Henize~2-10 
%\cut{(08$^{\rm h}$36$^{\rm m}$15$.\!\!^{\rm s}$20, -26$^\circ$24$^\prime$34$.\!\!^{\prime\prime}$0 [J2000])}
was observed in the
$^{13}$CO($J$=2-1) (220.399 GHz) and HCN($J$=1-0) (88.63 GHz) lines on December
2007 with the 30m Institut de Radioastronomie Millim{\'e}trique 
(IRAM\footnote{This publication is 
partially based on observations carried out with the IRAM 30m telescope. 
IRAM is supported by INSU/CNRS (France), MPG (Germany) and IGN (Spain).}) telescope
on Pico Veleta. 
A single position (08$^{\rm h}$36$^{\rm m}$15$.\!\!^{\rm s}$20, 
-26$^\circ$24$^\prime$34$.\!\!^{\prime\prime}$0 [J2000]) was observed.
The spectra have a typical beam size of
$\sim$11$^{\prime\prime}$ in the $^{13}$CO($J$=2-1) line and
$\sim$28$^{\prime\prime}$ in the HCN($J$=1-0) line.  Calibration, pointing and
focus checks were performed regularly. 
%\new{Typical pointing accuracy is about 2$^{\prime\prime}$.} 

The data processing was done with the 
CLASS\footnote{See http://www.iram.fr/IRAMFR/GILDAS for more information 
about the GILDAS softwares.} softwares.
%(Pety 2005). 
The data were converted into flux density units, using S/T$_{\rm mb}=4.95$~Jy/K
(Rohlfs \& Wilson 1996).
The $^{13}$CO($J$=2-1) and HCN($J$=1-0) spectra were smoothed to a velocity resolution of 5~km~s$^{-1}$ and 10~km~s$^{-1}$, 
respectively, and the final rms noise was 40~mJy and 5~mJy per channel, respectively.

\subsubsection{APEX Telescope}

We observed Henize~2-10 in the $^{13}$CO($J$=2-1) (220.399 GHz) with the
Atacama Pathfinder Experiment telescope (APEX\footnote{This publication is
partially based on data acquired with the Atacama Pathfinder Experiment (APEX, ESO project 081.F-9813).
APEX is a collaboration between the Max-Planck-Institut fur Radioastronomie,
the European Southern Observatory, and the Onsala Space Observatory.},
G{\"u}sten et al. 2006) on August 28th and 31st 2008 and on December 17th 2008, in the same 
position used for the IRAM-30m observations. The typical FWHM  beam size for the 12m antenna
at the $^{13}$CO($J$=2-1) frequency is 28$^{\prime\prime}$. The line was
measured in upper sideband (USB) using the APEX-1 facility receiver in
combination with the Fast Fourier Transform Spectrometer. 
%\new{Typical pointing accuracy is about 2$^{\prime\prime}$.} 

The data were reduced using the CLASS software. A first order polynomial 
was fitted to the line-free channels and subtracted off the baseline.
%\cut{To increase the
%signal to noise ratio, the spectrum was smoothed to a velocity resolution of
%%5.3 
%10.7 km~s$^{-1}$.}  
We have converted our line intensities into a flux density
scale, 
%(i.e. in Jansky units), 
assuming the aperture efficiency observationally
determined by G{\"u}sten et al. (2006): 1~K of antenna temperature corresponds
to 39 Jy. 
To increase the
signal to noise ratio, the spectrum was smoothed to a velocity resolution of
%5.3 
10.7 km~s$^{-1}$.
The noise level in our final spectrum is of the order of 70~mJy per velocity channel.

\subsubsection{SEST Telescope}

Henize~2-10 was %observed 
mapped in the $1.2$~mm continuum with the bolometer array SIMBA
at the SEST on the nights %6-7 
7-8 and 8-9 June 2002. The beam size of the SEST at this
wavelength is 21$^{\prime\prime}$. Background subtraction was obtained by fast scanning
(80$^{\prime\prime}$/sec) of the source through the array field of view in AZ and ALT,
the map size was 480$^{\prime\prime}\times 240^{\prime\prime}$. The atmophere opacity was
$\tau\sim0.1$ during the observations. We reduced the data with the
MOPSI (Zylka 1998) package. The total integration time was 4 hours and 50 minutes. 
The flux calibration was obtained by comparison with planet Uranus.
We detected an un-resolved source whose flux is 52$\pm$5~mJy. 
%The flux calibration was obtained by comparison with planet Uranus.

\subsection{Interferometric SMA observations}

Submillimeter interferometric observations of Henize~2-10 in the CO($J$=2-1)
(230.538 GHz) line were carried out on February and March 2008 with the
Submillimeter Array\footnote{The Submillimeter Array is a joint project between
the Smithsonian Astrophysical Observatory and the Academia Sinica Institute of
Astronomy and Astrophysics and is funded by the Smithsonian Institution and the
Academia Sinica.} (Ho et al. 2004).  We obtained data with the array in the
extended and compact configurations, half track each, and with six antennas and
seven antennas respectively.  The phase centre of the observations was the
same position used for the IRAM-30m observations.  The range of baselines of
our dataset spans from the shadowing limit to $\sim 220$~m.  The observations
cycled between Henize~2-10 and two gain calibrators (0826-225 and 0730-116),
with 15 minutes on the target and 7 minutes on the calibrators.  Passband
calibration was ensured observing 3C273 and 3C279.  0730-116 was used to set
the absolute flux scale. The flux of this calibrator at the time of
observations was determined using values from monitoring observations of the
SMA standards (moons and planets).  The SMA receivers operate in a double-side
band mode, with the upper and lower side bands separated by 8 GHz.  We tuned to
observe the CO(2-1) in the upper sideband and the $^{13}$CO(2-1) and
C$^{18}$O(2-1) in the lower sideband.

The initial flagging and calibration of the raw visibility data were done using
the IDL-based MIR
package\footnote{http://sma-www.cfa.harvard.edu/miriadWWW/manuals/\\
SMAuguide/smauserhtml/}.  The MIRIAD package was used for imaging and
deconvolution of the calibrated visibilities and for the data analysis.  The
continuum was constructed from the line free channels in the visibilitiy
domain.  Continuum and line maps from the combined datasets were made with
natural weighting, providing a final synthesized beam FWHM of
1$.\!\!^{\prime\prime}$9$\times$1$.\!\!^{\prime\prime}$3 (at a position angle
of 5~deg) for the line map, which corresponds to a physical size of
80$\times$60~pc for Henize~2-10, and of
1$.\!\!^{\prime\prime}$9$\times$1$.\!\!^{\prime\prime}$4 (at a position angle
of 6~deg) for the continuum map.  The rms noise is 19~mJy~beam$^{-1}$ per
5~km~s$^{-1}$ channel in the CO($J$=2-1) line data and 
%0.8
1~mJy~beam$^{-1}$ in the continuum.

%__________________________________________________________________

\section{Results}\label{results}
%\section{\bf{Observational} Results}\label{results}

\subsection{Millimeter Continuum }\label{mmcontinuum}

No millimeter continuum detection was obtained at the high angular resolution
of the SMA. 
%Even if some positive flux is seen in the continuum maps, no clear
%detection above a 5$\sigma$ upper limit of $\sim$5~mJy/beam was obtained.
We obtain no clear detection above a 5~$\sigma$ upper limit of 5~mJy/beam, which converts 
into a total gas+dust mass between 5~10$^6$~$M_{\odot}$ and 1.6~10$^7$~$M_{\odot}$, assuming 
a range of temperatures between 20~K and 50~K and a dust opacity of 0.005 cm$^2$~g$^{-1}$ 
(see discussion in section \ref{mma}).
%and a gas-to-dust ratio of 150.

Our SIMBA single dish 1.2~mm flux is consistent with the MAMBO 1.3~mm 
flux of $\sim (56 \pm 14)$~mJy reported by Galliano et al. (2005),
while, integrating the SMA continuum map over the 30m beam, we recover 
%at most 
17~mJy with the interferometer, which is the 30\% of the single dish flux.
%which is significantly stronger than our 1.3~mm SMA value. 
The most likely explanation, given also the extent ($\sim$15$^{\prime\prime}$) of the 450 and 850~$\mu$m
emission detected 
by Galliano et al. (2005), is that 
our interferometric observations filter out the most extended emission.
%, which is seen 
%at the 11$^{\prime\prime}$ resolution of the MAMBO observations 
%with MAMBO and is seen also in the SCUBA observations at 450 and 850 $\mu$m by Galliano et al. (1994).
To estimate this effect we used the approach outlined by Wilner \& Welch (1994). Given a 
minimum baseline length, their formalism allows us to estimate the fraction of flux seen by the interferometer 
for a Gaussian source with a given FWHM. For our observations, the minimum projected baseline
of 7~k$\lambda$, where $\lambda$ is the wavelength of our observations,
allows us to recover 100\% of the total flux %density 
for sources comparable to the beamsize, but
%$<40-77$\% 
only $\sim$40\% of the total flux density for a source of FWHM %$\gtrsim8-15$ arcsecs 
15$^{\prime\prime}$
(650~pc at a distance of 9~Mpc).
%$<77-90$\% of the total flux density for sources with FWHM $\gtrsim5-8$ arcsecs (220-350~pc at a distance of 9~Mpc).}
%\new{Wilner \& Welch (1994) quantify the amount of flux density one can expect to recover 
%from objects with extended structure using an interferometer. Using the same formalism, if the extended structure of the source
%can be described by a Gaussian with FWHM = 1$.\!\!^{\prime\prime}$8, we estimated
%that almost 100\% of the total flux density 
%is recovered by the SMA interferometer at 1.3~mm. In contrast, only 
%%70\% of the total flux density of a 9$^{\prime\prime}$ ($5*beam$) gaussian source 
%3\% of the total flux density of a $30^{\prime\prime}$ ($= \lambda/S_{min}$, where $S_{min}$ is the minimum baseline length 
%projected on the sky, which is equal to 7~k$\lambda$ for our SMA observations)
%gaussian source is present in the interferometric map at 1.3~mm.} 
%\cut{Indeed, a simple estimate, based on our shortest baselines and using the 
%formalism of Wilner \& Welch~(1994), shows that
%we do expect to filter out
%most of the emission more extended than 5-8~arcsec (220-350~pc at a distance of 9~Mpc).}

%\subsubsection{Comparison with 3.6~cm continuum and HST}\label{cont}

\subsection{Molecular Gas Morphology}\label{morphology}

\subsubsection{Single dish observations}

\begin{figure}
\centering
\includegraphics[scale=0.47, angle=-90]{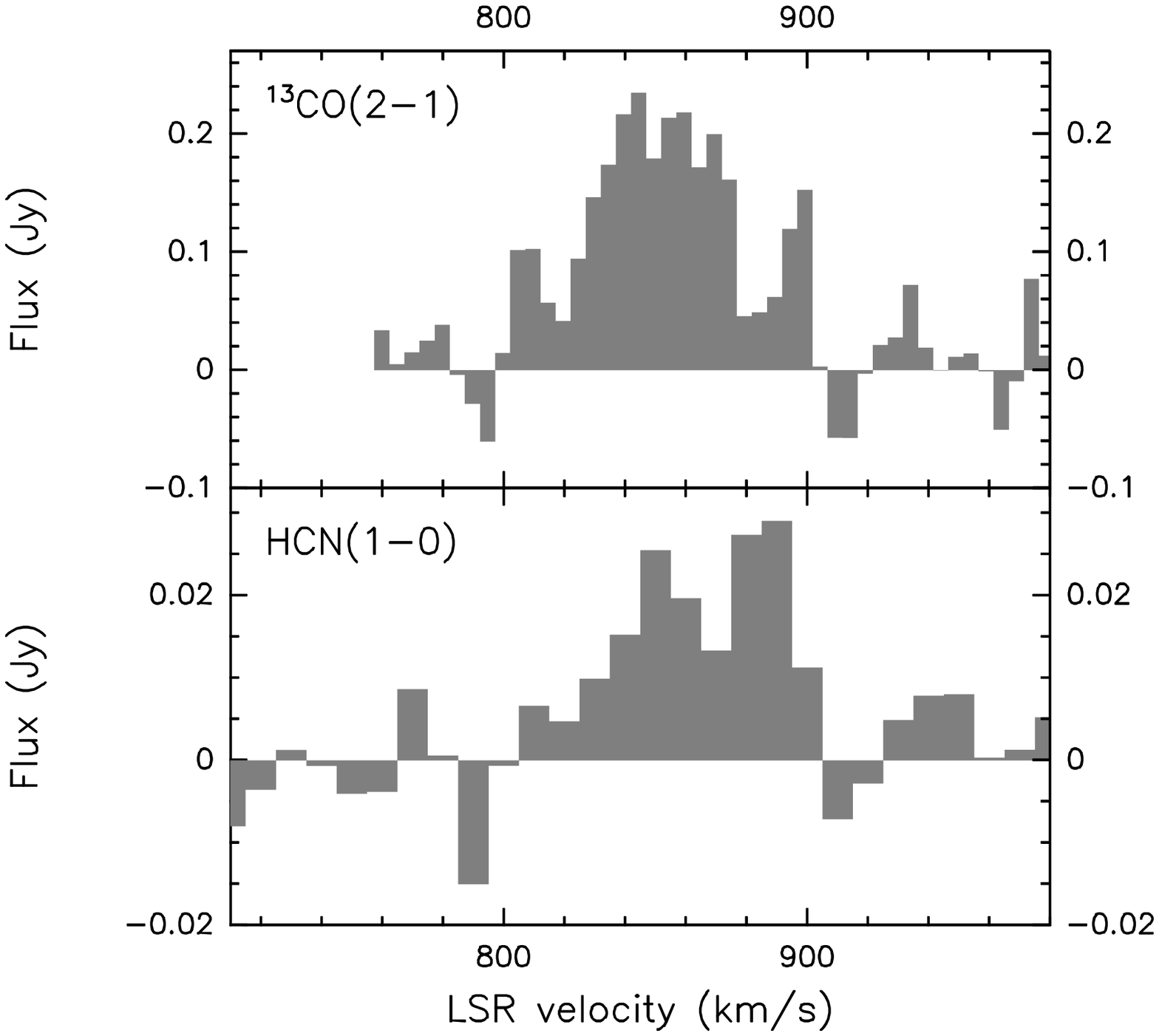}
\caption{$^{13}$CO($J$=2-1) (top) and HCN($J$=1-0) (bottom) IRAM-30m spectra.}
\label{30m}
\end{figure}

In Fig.~\ref{30m}, we show the observed IRAM-30m spectra of the
$^{13}$CO($J$=2-1) and HCN($J$=1-0) transitions.
Both lines are detected 
%(as was aspected for this galaxy, which is very rich in molecular gas) 
and they are approximately 100 km~s$^{-1}$ wide, 
consistent with previous observations of CO and isotopologues by Kobulnicky et al. (1995).

The line parameters have been estimated fitting a Gaussian profile to each
spectrum.
Table \ref{IRAMeAPEX_lines} summarizes the beam size at the observed frequencies, 
the peak and integrated flux densties and the line width of the observed emission lines.

The HCN(1-0) total intensity that we derive from our spectrum
($\le$2~Jy\,km\,s$^{-1}$) is significantly lower than the value
found by Imanishi et al.~(2007) in their NMA integrated map 
($\sim$6~Jy\,km\,s$^{-1}$). This is surprising as the single dish 
telescope should have detected the same or more emission than the
interferometer. Both the IRAM-30m and the NMA detections do not have
a very high signal to noise ratio and higher sensitivity observations
will be needed in the future to confirm the flux of this source.
We cannot exclude the possibility of a problem with the calibration of the 
IRAM spectrum. %\cut{as discussed below}.
\begin{table*}
\centering
%\begin{minipage}[t]{\columnwidth}      %per fare una tabella con footnotes
\caption{IRAM-30m and APEX $^{13}$CO($J$=2-1) and HCN($J$=1-0) emission lines
parameters.}\label{IRAMeAPEX_lines}
\smallskip 
\begin{threeparttable}
\renewcommand{\footnoterule}{}  % to avoid a line before footnotes
\begin{tabular}{lccccc}
\hline \hline
Line & Beam Size & Flux$_{peak}$ & FWHM & V$_{\rm LSR}$ & Integrated Flux\\
~&[arcsec] & [Jy] & [km~s$^{-1}$] & [km~s$^{-1}$] & [Jy km~s$^{-1}$]\\
(1)&(2)&(3)&(4)&(5)&(6)\\
\hline
IRAM-30m & ~ & ~ & ~\\
$^{13}$CO($J$=2-1) & 11 & 0.21 & 56 & 850 & 13$\pm$1 \\
HCN($J$=1-0) & 28  & 0.025 & 64 & 860 & 2.0$\pm$0.5 \\
%\hline
~&&&\\
APEX & ~ & ~ & ~\\
$^{13}$CO($J$=2-1) & 28 & 0.5 & 42 & 855 & 22$\pm$2 \\
\hline
\end{tabular}
%\begin{tablenotes}
%\item[] NOTES. -- Col.(1): Observed emission line. Col.(2): Beam size. Col.(3): Peak flux density of the emission. Col.(4): Line width of the gaussian fit of the emission at FWHM. Col.(5): Peak velocity of the emission line. Col.(6): Integrated flux density of the emission.
%\end{tablenotes}
\end{threeparttable}
%\end{minipage}
\end{table*}

\begin{figure}
\centering
\includegraphics[scale=0.55, angle=-90]{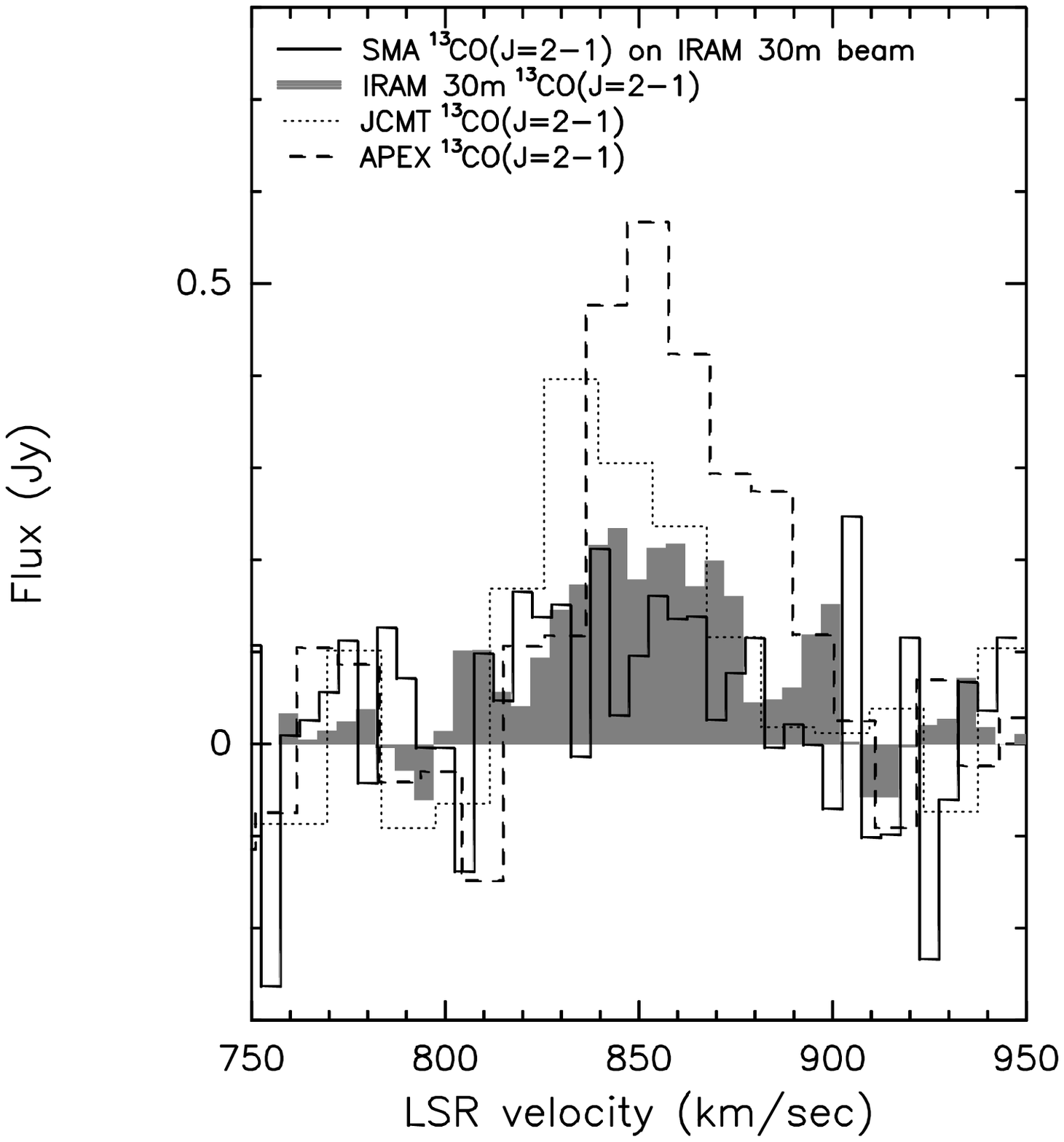}              %prova_fig4INO.eps
\caption{Comparison between the $^{13}$CO($J$=2-1) spectra from IRAM-30m
(grey shaded), SMA (solid line, integrated over the IRAM-30m beam area),
JCMT (dotted line; Baas et al. 1994) and APEX (dashed line).}
\label{13coSMAeIRAM}
\end{figure}

In Fig.~\ref{13coSMAeIRAM} we show the
observed APEX $^{13}$CO($J$=2-1) spectrum (dashed line).
The parameters of the emission line from a Gaussian fit to the spectrum 
are presented in Table \ref{IRAMeAPEX_lines}. In the same figure we show also
the IRAM-30m spectrum (grey shaded) and the JCMT spectrum from Baas et al.~(1994, dotted line).
Figure \ref{13coSMAeIRAM} shows that the IRAM-30m $^{13}$CO($J$=2-1) spectrum
is fainter than both the JCMT and the APEX spectra.
This could be due to either a problem in the calibration or pointing 
of the IRAM-30m data,
because of the large airmass of this southern object, or to the fact that the
$^{13}$CO($J$=2-1) emission is more extended than the IRAM-30m beam
(11$^{\prime\prime}$ at this frequency). Therefore part of the emission which
is seen by APEX and JCMT, with beam sizes of 28$^{\prime\prime}$
and 21$^{\prime\prime}$ respectively, is outside the IRAM beam. This explanation
is consistent with the SCUBA continuum maps of Galliano et al.~(2005), 
%\cut{who resolve the emission \cut{(see \S\ref{mmcontinuum})},} 
the non-detection of the
$^{13}$CO($J$=2-1) emission with the SMA (see below) and the detection of an extended 
component in the CO($J$=3-2) emission by Vanzi et al. (2009).

%\subsubsection{SMA observations}
\subsubsection{Comparison between single dish and interferometer}

\begin{figure}
\centering
\includegraphics[scale=0.55, angle=-90]{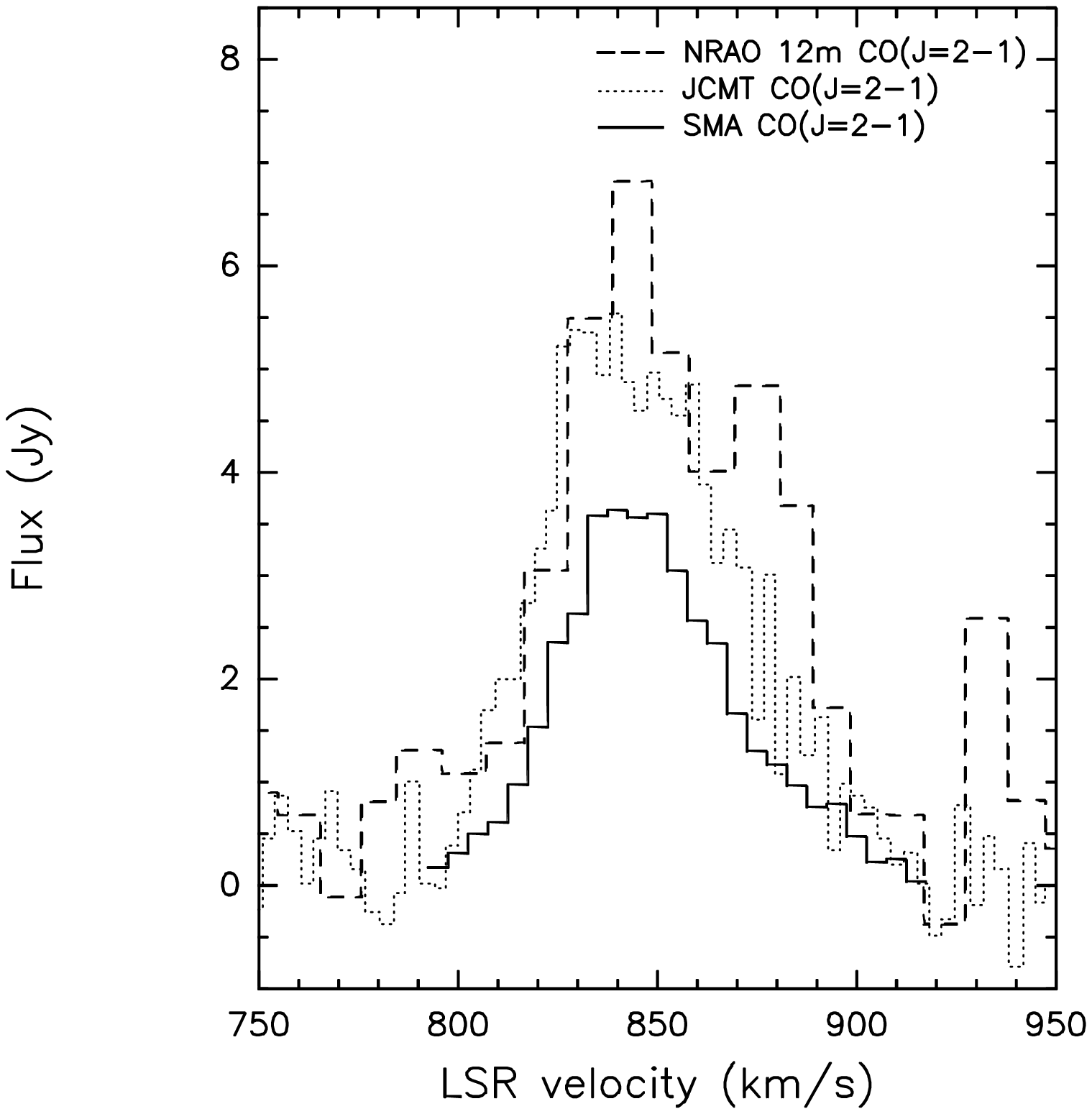}
\caption{Comparison between the NRAO-12m CO($J$=2-1) spectrum (dashed line) from Kobulnicky et 
al. (1995), our SMA CO($J$=2-1) spectrum (solid line), integrated over the whole region of the emission,
%integrated on the NRAO 12~m beam at the CO($J$=2-1) frequency (27$^{\prime\prime}$)
and the JCMT CO($J$=2-1) spectrum (dotted line) from Baas et al. (1994).}
\label{recovering}
\end{figure}

Figure \ref{recovering} shows the NRAO CO($J$=2-1) spectrum (dashed line)
from Kobulnicky et al. (1995) overlaid on our SMA integrated spectrum (solid line)
%integrated on the NRAO~12m beam size (27$^{\prime\prime}$), centred on the position $\alpha_{2000} = 08^h36^m15.\!\!^s17$ and $\delta = -26^{\circ}24^{\prime}34.\!\!^{\prime\prime}0$
and the %\cut{the} 
JCMT CO($J$=2-1) spectrum from Baas et al. (1994, dotted line).
The comparison shows that we recover 44\% of the single dish flux density, while the remaining part, which 
most probably comes from extended emission, is filtered out by the interferometer.
%at the higher SMA resolution.

Kobulnicky et al. (1995) found an elongated cloud of molecular gas
with a tail towards the south-east.
This tail, revealed in the OVRO CO($J$=1-0) map in a 
velocity interval between 805 and 835 km~s$^{-1}$, is not seen in our SMA CO($J$=2-1) data 
(with the exception of two 3~$\sigma$ level peaks), most likely because the higher resolution SMA observations 
filter out this extended emission. 
%The tail is seen at low velocities in the NRAO 12m spectrum 
%(Kobulnicky et al. 1995), 
%shown in Fig. \ref{recovering}, where it differs from our SMA spectrum. 
%%The NRAO 27$^{\prime\prime}$ beam filters out part of the emission, which is partly in the external region 
%%of the beam and partly out of the NRAO beam. 
%The flux density that is missing with our SMA observations 
%in the velocity interval of around 870-890 km/s is probably due to extended emission, 
%which is partially seen in the 
%low resolution ($6.\!\!^{\prime\prime}5 \times 5.\!\!^{\prime\prime}5$) 
%OVRO observations from Kobulnicky et a. (1995).
%and is missing with our higher resolution SMA observations.

As mentioned before, the correlator frequency coverage and configuration were
such that also $^{13}$CO($J$=2-1) and C$^{18}$O($J$=2-1) transitions were
observed.  From our IRAM-30m observations of $^{13}$CO($J$=2-1) we expect a
peak flux density of about 0.2 Jy on a beam size of around 11$^{\prime\prime}$.
The spectrum of our SMA $^{13}$CO($J$=2-1) observations integrated over the IRAM beam
is shown %in solid line 
in Fig. \ref{13coSMAeIRAM}, overlaid on the
IRAM-30m %grey shaded 
$^{13}$CO($J$=2-1) spectrum, the JCMT $^{13}$CO($J$=2-1)
spectrum from Baas et al. (1994) %in dotted line 
and the APEX $^{13}$CO($J$=2-1)
spectrum. %in dashed line.
%\begin{figure}
%\centering
%\includegraphics[sale=0.5, angle=-90]{13coSMA-IRAM-APEX-baasINO.eps}              %prova_fig4INO.eps
%\caption{Comparison between the $^{13}$CO($J$=2-1) spectra from IRAM-30m
%(grey shaded), SMA (solid line, integrated over the IRAM-30m beam area),
%JCMT (dotted line; Baas et al. 1994) and APEX (dashed line).}
%\label{13coSMAeIRAM}
%\end{figure}
%The superposition shows that we are missing ...\% of the single-dish flux coming from extended emission.
The rms of the SMA spectrum (roughly 0.1 Jy) is about half of the peak flux
%density 
expected from the single dish IRAM observations and does not permit a credible detection. 
%to detect the emission line with high enough signal to noise ratio.
%In fact, if we invert the 
%brightness temperature from the 30m-IRAM observations of $^{13}$CO($J$=2-1) (see  
%Figure \ref{30m}), 
%we obtain a flux density estimate of $\sim$25 mJy/beam, which is 
%under the noise level of our SMA observations. 
The C$^{18}$O($J$=2-1) %line 
emission, as one might expect, was detected neither in the APEX spectrum
%and, as expected, we were also not able to detect it with 
nor by the SMA.
%expected to be

The non-detection of the $^{13}$CO(2-1) line and the large fraction of the CO(2-1)
flux that is not recovered by our SMA observations are consistent with the 
presence of extended molecular emission to which the SMA is not sensitive, as discussed in the previous
paragraph, and with the fact that the SMA is only recovering a fraction of the
millimeter continuum emission. 

\subsubsection{Morphology of the molecular gas emission}

\begin{figure*}
\centering
\includegraphics[scale=0.92, angle=-90]{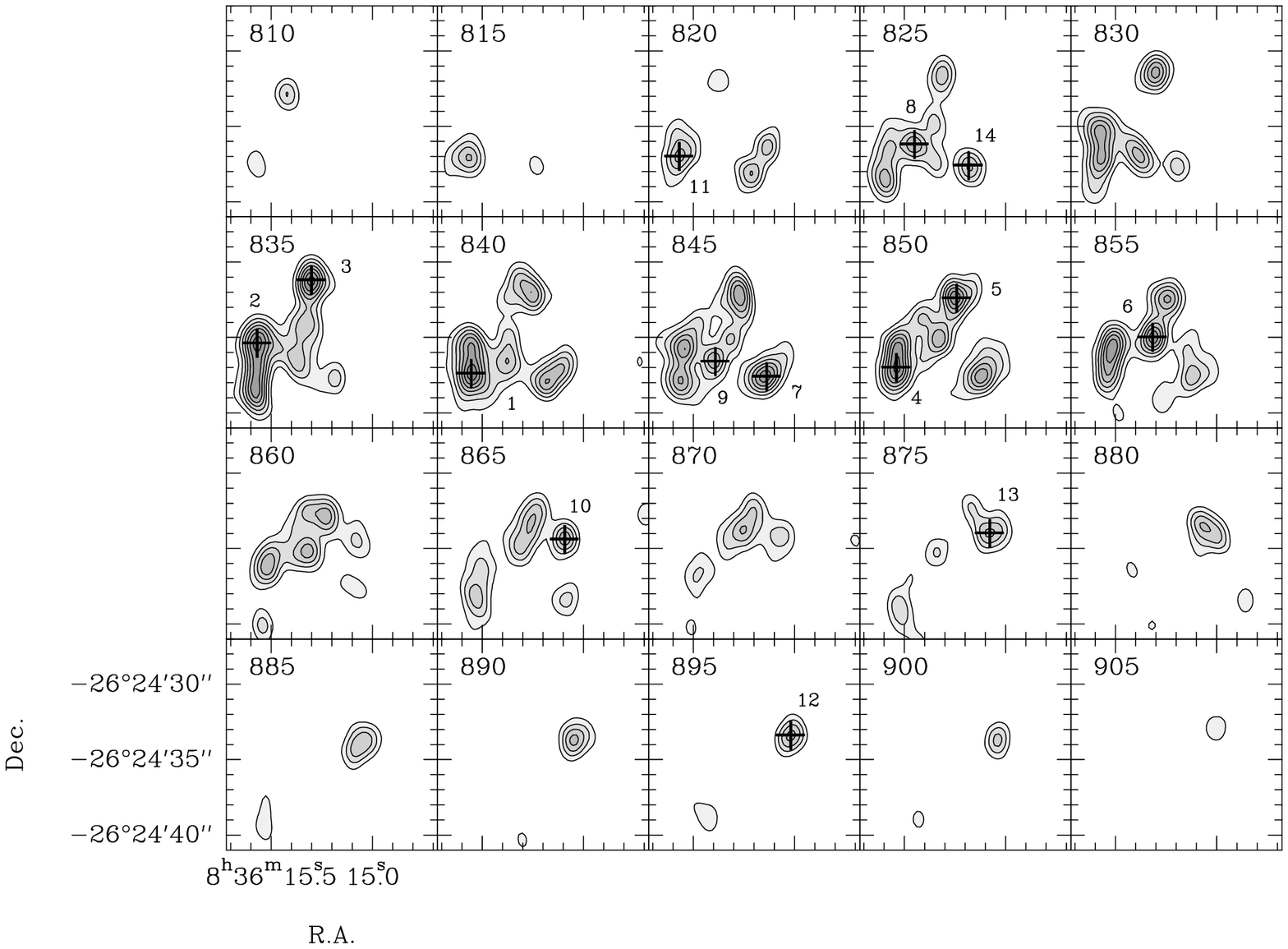}
\caption{Channel map of the CO($J$=2-1) line. The contour levels are from
5~$\sigma$ ($\sigma$=19~mJy/beam) in steps of 3~$\sigma$. The black numbered
crosses indicate the peak position (in the channel corresponding to the
peak velocity) of the emission of the 14 molecular clouds identified by
the Clumpfind algorithm (see \S\ref{clumpfind} and Table \ref{tab_param}).}
\label{canali}
\end{figure*}

Our SMA spectral line data resolve the OVRO CO($J$=1-0) emission from
Kobulnicky et al. (1995) into several compact sources. The velocity structure of
the CO($J$=2-1) emission is given in Fig.~\ref{canali}, which shows the channel map of
the CO($J$=2-1) emission.
The map reveals the complex velocity structure of the CO emission, with most of
the emission centred on an LSR velocity around 850 km s$^{-1}$. Several
velocity components can be distinguished in the emission and some of them are
comparable with the beam size. The CO components are well separated in both
position and velocity and spread over a velocity interval of about 100 km
s$^{-1}$, as already revealed from the 
%IRAM-30~m 
single dish observations of
$^{13}$CO($J$=2-1) and HCN($J$=1-0) (see Fig.~\ref{30m} and Fig.~\ref{13coSMAeIRAM}).

%\subsubsection{Comparison with CO(1-0) di Kobulnicky et al. 1995}\label{kobuln}

%Kobulnicky et al. (1995) found an elongated cloud of molecular gas with a 
%tail towards the south-east.
%Our SMA spectral line data resolve the OVRO CO($J$=1-0) emission from Kobulnicky et al. (1995) 
%in several compact sources. 

%\subsubsection{Comparison with HST images}\label{hst}

%Figure \ref{co-hst} gives the velocity-integrated CO($J$=2-1) intensity map 
%in contours, overlaid on the HST V-band map (\emph{Top Panel}) and on the HST H$\alpha$ map, 
%showing that some of the CO clouds are indeed associated with the SSCs.
%(\emph{Bottom Panel})
%\begin{figure}
%\centering
%% \includegraphics[angle=-90,scale=0.4]{co_i.eps}
%% \includegraphics[angle=-90,scale=0.4]{co_halfa.eps}
%\caption{\emph{Top Panel}: Line integrated CO($J$=2-1) map (\emph{contours}), 
%overlaid on the HST V-band map. \emph{Bottom Panel}: Line integrated CO($J$=2-1) 
%map (\emph{contours}), overlaid on the HST H$\alpha$ map. The contours are 
%from 5~$\sigma$ ($\sigma$=) in steps of 1~$\sigma$.}
%\label{co-hst}
%\end{figure}

%comparison with HST...????
%clump dietro all'HST lungo la linea di vista?...v. david...

%\subsubsection{Comparison with 3.6cm continuum}

\begin{figure}
\centering
\includegraphics[angle=-90, scale=0.4]{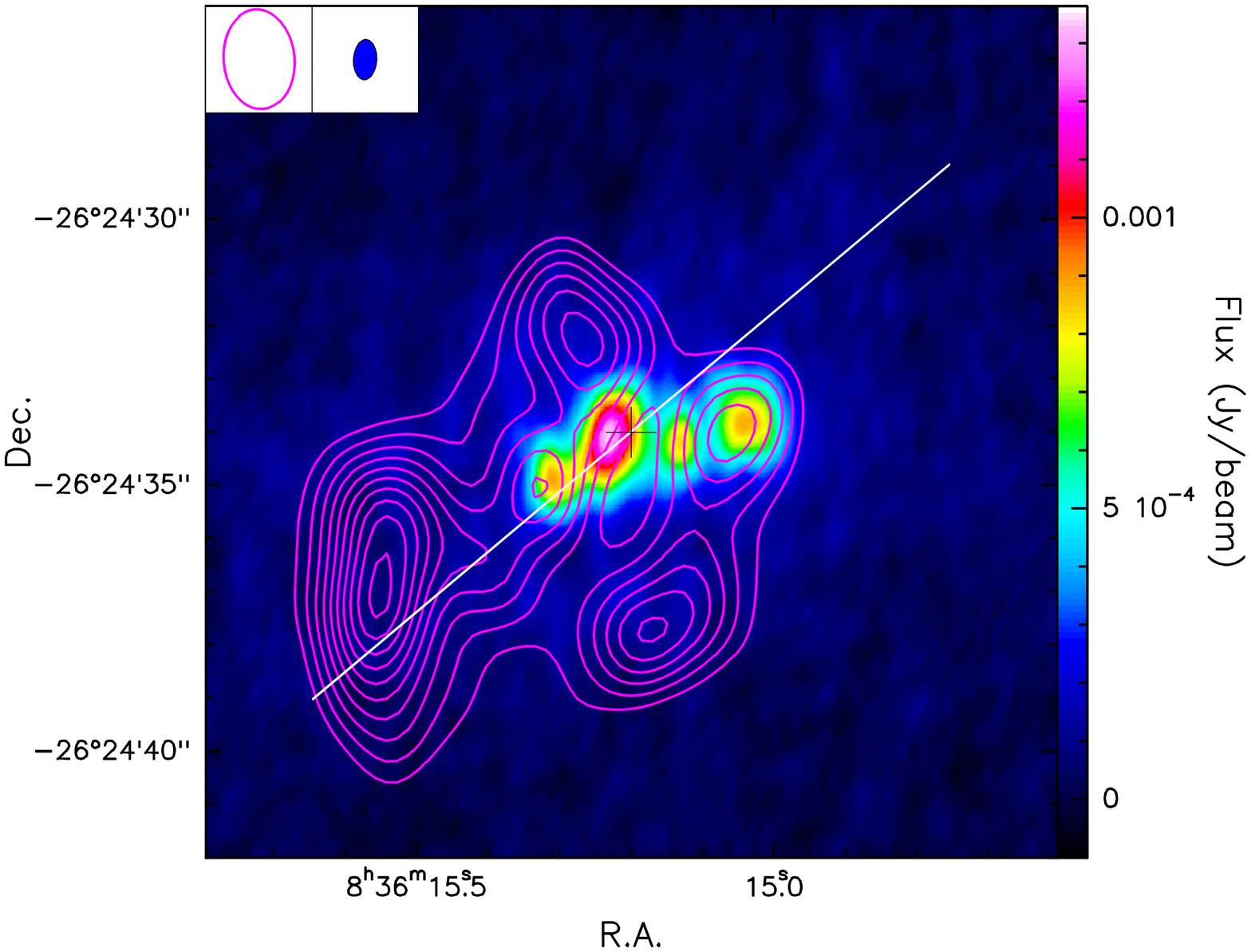}
%\hspace{2em}
%\includegraphics[angle=-90, scale=0.4]{bla_1e3mm_colorINO.eps}
%\includegraphics[scale=0.4]{imanishi_cut.ps}
\caption{SMA integrated CO($J$=2-1) map (contours),
overlaid on the VLA 3.6~cm continuum image from Johnson \& Kobulnicky (2003).
The contour levels are from 5~$\sigma$ in steps of 2~$\sigma$ ($\sigma$=0.8 Jy
beam$^{-1}$ km s$^{-1}$). The white line and the cross indicate respectively
the orientation of the position-velocity cut, shown in Fig.\ref{posvel} (see
\S\ref{dinamica}), and the central position of the map.
%\emph{Right}:
%SMA integrated CO($J$=2-1) map (\emph{contours}), overlaid on the SMA 1.3~mm
%continuum map. The color scale shown on the right side of each map is in
%Jy/beam. 
}
\label{co_cont}
\end{figure}

Figure \ref{co_cont} shows the SMA CO($J$=2-1) velocity-integrated intensity map 
in solid contours overlaid on the VLA 3.6~cm continuum 
(Johnson \& Kobulnicky 2003). 
%and on our SMA 1.3~mm continuum (right panel).
Part of the centimeter continuum emission, associated with the super star clusters, is coincident with the 
most compact molecular gas sources in the CO($J$=2-1) emission line. In particular, the centimeter sources 
1, 2 and 5 (see Johnson \& Kobulnicky 2003)
%Fig.\ref{continui}) 
correspond respectively to our sources 12, 13 and 6 
(see Fig.~\ref{canali} and Table~\ref{tab_param}). 
Some of the CO emission is not associated with the SSCs.
%and may correspond to less dense molecular gas not involved in the SSC formation.

The strongest 3.6~cm source, which does not correspond to any significant amount of molecular emission,
is likely in an older evolutionary stage, having already dispersed most of the surrounding molecular 
material from which it formed.
%Besides, the opacity of the CO line
%is optically thick and this 
%has the effect to trace only the surface of the emission, 
%corresponding to an optical depth equal to 1 (?). 
%while the hot CO clouds might be hidden by foreground colder CO.

%It is interesting to note that the SMA 1.3~mm continuum sources, which are likely associated to the dust emission, 
%do not correspond to any strong molecular gas emission.
%, although there might be an effect due to the opacity 
%of the CO emission,
%It is possible that this is due to the opacity of the CO line,
%which prevents from seeing molecular gas emission peaks towards the dense 
%hot inner regions traced by the mm continuum.

We performed Gaussian fits to the spectra of the lines observed with SMA,
integrated over the whole region of the emission, to estimate the peak and
integrated flux density and the line width of the observed emission lines.
Table \ref{SMA_lines} summarizes the fitting parameters.
\begin{table}
\centering
%\begin{minipage}[t]{\columnwidth}      %per fare una tabella con footnotes
\caption{Integrated SMA CO($J$=2-1) and $^{13}$CO($J$=2-1) lines
parameters.}\label{SMA_lines}
\smallskip 
\begin{threeparttable}
\renewcommand{\footnoterule}{}  % to avoid a line before footnotes
\begin{tabular}{ccccc}
\hline \hline
Line & Flux$_{peak}$ & FWHM & V$_{\rm LSR}$ & Integrated Flux\\
~&[Jy] & [km~s$^{-1}$] & [km~s$^{-1}$] & [Jy km~s$^{-1}$]\\
(1)&(2)&(3)&(4)\\
\hline
$^{12}$CO($J$=2-1) & 3.5 & 48 & 846 & 178.4$\pm$1.6\\
$^{13}$CO($J$=2-1) & $\lesssim 0.3$ & -- & -- & $\lesssim 12$\\
\hline
\end{tabular}
%\begin{tablenotes}
%\item[] NOTES. -- Col.(1): Observed emission line. Col.(2): Peak flux density of the emission. Col.(3): Integrated flux density of the emission. Col.(4): Line width of the gaussian fit of the emission at FWHM.
%\item[a] The flux densities are upper limits because of the low signal to noise ratio of the $^{13}$CO($J$=2-1) spectrum.
%\end{tablenotes}
\end{threeparttable}
%\end{minipage}
\end{table}
The values of the peak and integrated flux densities of the $^{13}$CO($J$=2-1)
emission line, observed with SMA, are upper limits because the line is not detected.
%because of the low signal to noise ratio of the spectrum.  
It is thus possible to derive a lower limit to
the flux density ratio between the CO($J$=2-1) and $^{13}$CO($J$=2-1) emission
lines. We estimate a CO($J$=2-1)/$^{13}$CO($J$=2-1) ratio $\gtrsim 14.9$.

%___________________________________________________________________

\section{The structure of the CO($J$=2-1) emission}\label{derivation}

\subsection{Identification of the clouds: Clumpfind}\label{clumpfind}

We applied the Clumpfind algorithm (\cite{williams94}) to the SMA CO($J$=2-1) data
cube to decompose the emission into discrete structures and identify the
individual molecular clouds. 
Clumpfind works by making a contour map through the data cube to locate the peak 
flux densities and then steps down the contour levels to determine where the clumps are 
located. Clumpfind allows the user to adjust the contour spacing and the lower contour level.
The advantage of this algorithm is that we make
no assumptions about the cloud shape. The specific intensity contour used as
emission threshold, which defines the borders (boundaries) of the individual
identified clouds, is 0.1 Jy/beam, corresponding to 5~$\sigma$ level.
%($\sigma$=20 mJy/beam). 
The step level used to disantangle the emission and
identify isolated peaks is 60 mJy/beam (3~$\sigma$).
To be sure the algorithm is not introducing artefacts, we decided to 
limit the analysis to the clouds with peak flux density larger than 
15~$\sigma$.
In this way, we identified 14 separate CO clouds. 

%There are evidences of double
%peak emission in the Clumpfind analysis and it is noteworthy 
Is is important to realize that the Clumpfind
algorithm does not take into account effects of self--absorption, 
which manifest in a double peak line profile:
%\cut{but} 
it would then try to separate the two peaks of a single self--absorbed clump
spectrum 
%of a single clump
in two distinct clouds 
%\new{a single clump with} 
%\cut{double peak emission}
with same spatial position and different velocity.
%\new{spectrum}. 
With our data we do not see a clear evidence that this is effecting a significant 
fraction of the clumps we identify, as most of the clumps are also spatially distinct. 

To emphasize the
complex substructure of the emission we marked with black crosses in Fig.
\ref{canali} (the channel map of the emission) the position of the clouds
identified by the Clumpfind algorithm, in the channel relative to the peak of
each cloud emission. The 14 components are well visible in the plot and are
labeled by their respective number.

The parameters for our final cloud decomposition of the data set, given by
Clumpfind, are presented in Table \ref{tab_param}. The angular sizes have been
computed from the 5~$\sigma$ contour levels and deconvolved with a simple
Gaussian deconvolution for the beam size, assuming source and beam to be
Gaussian. The velocity widths have been deconvolved as well for the velocity
resolution. %\cut{The size of some of the clouds is comparable to our beam,
%suggesting that part of the CO emission is not resolved at a linear resolution
%of $\sim$80~pc.} 
Table \ref{tab_param} gives also the derived masses of the
identified clouds (see next section for the details).

%Figure \ref{clump} shows some (?) of the clouds identified by Clumpfind...
%\begin{figure}
%\centering
%% \includegraphics[angle=-90, scale=0.4]{}
%\caption{}
%\label{clump}
%\end{figure}

\subsection{Derivation of Physical Parameters of the Clouds}\label{parameters}

Several methods can be used to obtain molecular mass estimates.  The molecular
hydrogen column density, N(\rm H$_2$), can be estimated from the CO flux,
$S_{\rm CO}$, using a CO-to-H$_2$ conversion factor, X$_{\rm CO}$ (X$_{\rm
CO}$=N(H$_2$)/I$_{\rm CO}$), which is empirically determined from Galactic
molecular clouds.  It is a matter of debate what is the best choice for X$_{\rm
CO}$.  Dwarf galaxies, possibly due to their low metallicities, seem to have a higher
standard conversion factor than large metal-rich spirals (Maloney \& Black
1988; Verter \& Hodge 1995; Wilson 1995; Arimoto et al. 1996).  However,
Henize~2-10 is a dwarf galaxy with roughly solar metallicity (12+log(O/H)=8.93;
Vacca \& Conti 1992). Moreover, Blitz et al. (2007), studying
extragalactic Giant Molecular Clouds (GMCs), conclude that there is no clear
trend of X$_{\rm CO}$ with metallicity.  We thus estimate the molecular mass
($M_{\rm mol}$) of the clouds using the Galactic value for the CO-to-H$_2$
conversion factor, X$_{\rm COgal}=(3 \pm 1) \times 10^{20}$
cm$^{-2}$~(K~km~s$^{-1}$)$^{-1}$ (Strong et al. 1988; Sanders et al. 1987)
which also includes a factor of 1.36 to account for the helium mass
contribution. Following Wilson \& Scoville~(1990) we use:
\begin{equation}
 M_{\rm mol}=1.61 \times 10^{4} \,\left({115\,\rm GHz}\over{\nu}\right)^2 \,d_{\rm Mpc}^2 \,{{ S_{\rm CO}}\over{ R_{\rm 21}}} \,\,\, M_{\odot}
\label{mmol}
\end{equation}
where $S_{\rm CO}$ is the CO($J$=2-1) flux in Jy~km~s$^{-1}$, 
$d_{\rm Mpc}$ is the distance and $R_{\rm 21}$ is the CO($J$=2-1)/CO($J$=1-0)
line ratio, which we assumed equal to 0.89 (\cite{braine93}). 
The latter value is also consistent with the range of ratios derived for this 
galaxy (Meier et al. 2001; Baas et al. 1994) and for dwarf irregular galaxies (Petitpas \& Wilson 1998).

The masses of the individual clouds can also be calculated from the measured
sizes and velocity dispersions using the virial theorem.
The underlying assumption is that the internal kinetic energy and the
gravitational energy are in equilibrium (\cite{solomon87}).
Assuming the source to be spherical and homogeneous and neglecting 
contributions from magnetic field and surface pressure, the virial
cloud mass is given by:
\begin{equation}
 M_{\rm VIR}=0.509 \,d(\rm kpc) \,\Theta_{\rm S}(\rm arcsec) \,\Delta \rm V^2_{1/2}(\rm km \,s^{-1}) \,\,\, M_{\odot}
\label{virial}
\end{equation}
(MacLaren et al. 1988), where $\Delta {\rm V_{1/2}}$ is the measured full-width
velocity at half-maximum intensity in km~s$^{-1}$ and $\Theta_{\rm S}$ is the
cloud angular diameter in arcsec.

The masses derived for each cloud by means of the two different methods
described above are presented in Table \ref{tab_param}.
\begin{table*}
\centering
\caption{Physical parameters of the SMA CO($J$=2-1) molecular clouds
identified with Clumpfind.
}\label{tab_param}
\smallskip 
\begin{threeparttable}
\renewcommand{\footnoterule}{}  % to avoid a line before footnotes
\begin{tabular}{cccccccccccc}
\hline \hline
N & \multicolumn{2}{c}{Peak Position} &V$_{\rm peak}$ &S$_{\rm peak}$ &Radius\tnote{a} &$\Delta{\rm V}_{1/2}\tnote{a}$ &S$_{\rm CO(J=\rm 2-1)}$ &$M_{\rm VIR}$  &$M_{\rm mol}$ & $\Sigma_{\rm VIR}$ & $\Sigma_{\rm mol}$\\
%per calcolare sigma: con la massa viriale ff, ma c'era anche quella di blitz
~&R.A.[J2000]&Decl.[J2000]&[km/s] &[Jy/beam] &[arcsec]& [km/s]& [Jy~km/s]& [$\times 10^6\,M_{\odot}$]& [$\times 10^6\, M_{\odot}$]& [g/cm$^{2}$] & [g/cm$^{2}$]\\
(1)&(2)&(3)&(4)&(5)&(6)&(7)&(8)&(9)&(10)&(11)&(12)\\
\hline
1  &  8:36:15.53 & -26:24:37.4 & 840 & 0.48 & 1.3 &  18 &  10.7 & 3.6 & 3.9 & 0.08 & 0.09 \\
2  &  8:36:15.54 & -26:24:35.4 & 835 & 0.47 & 1.2 &  25 &  13.4 & 6.9 & 4.9 & 0.16 & 0.11 \\
3  &  8:36:15.30 & -26:24:31.2 & 835 & 0.46 & 1.2 &  18 &   7.9 & 3.3 & 2.9 & 0.09 & 0.08 \\
4  &  8:36:15.52 & -26:24:37.0 & 850 & 0.46 & 0.8 &  19 &   7.5 & 2.9 & 2.7 & 0.14 & 0.14 \\
5  &  8:36:15.24 & -26:24:32.4 & 850 & 0.43 & 1.2 &  25 &  12.7 & 7.1 & 4.6 & 0.17 & 0.11 \\
6  &  8:36:15.31 & -26:24:35.0 & 855 & 0.42 & 1.0 &  20 &   8.3 & 3.9 & 3.0 & 0.13 & 0.10 \\
7  &  8:36:15.14 & -26:24:37.6 & 845 & 0.42 & 1.2 &  13 &   8.1 & 1.9 & 3.0 & 0.05 & 0.07 \\
8  &  8:36:15.43 & -26:24:36.2 & 825 & 0.33 & 0.8 &  20 &   5.0 & 3.0 & 1.8 & 0.16 & 0.09 \\
9  &  8:36:15.38 & -26:24:36.6 & 845 & 0.32 & 1.2 &  26 &   8.1 & 7.4 & 3.0 & 0.19 & 0.07 \\
10 &  8:36:15.10 & -26:24:34.4 & 865 & 0.31 & 1.1 &  20 &   4.1 & 4.0 & 1.5 & 0.12 & 0.05 \\
11 &  8:36:15.54 & -26:24:37.0 & 820 & 0.29 & 1.0 &  15 &   5.0 & 2.1 & 1.8 & 0.08 & 0.06 \\
12 &  8:36:15.03 & -26:24:33.4 & 895 & 0.29 & 1.0 &  19 &   5.2 & 3.1 & 1.9 & 0.11 & 0.07 \\
13 &  8:36:15.09 & -26:24:34.0 & 875 & 0.28 & 0.9 &  19 &   5.0 & 3.1 & 1.8 & 0.12 & 0.07 \\
14 &  8:36:15.19 & -26:24:37.6 & 825 & 0.28 & 1.2 &  24 &   5.3 & 6.5 & 1.9 & 0.16 & 0.05 \\
\hline
\end{tabular}
\begin{tablenotes}
\item[] NOTES. -- Col.(1): Number of the cloud. Col.(2)-(3): Peak position of the emission. 
Col.(4): Peak velocity of the emission. Col.(5): Peak flux density of the emission. Col.(6): Deconvolved radius 
(at 5~$\sigma$ contour level) of the cloud. Col.(7): Deconvolved FWHP line width.
Col.(8): Integrated flux. Col.(9): Virial mass. Col.(10): 
Molecular gas mass from SMA CO($J$=2-1).
Col.(11): Surface density of the clouds, computed 
from the virial masses (Col.9). Col.(12): Surface density of the clouds, computed from the molecular 
gas masses (Col.10).
\item[a] Deconvolved.
\end{tablenotes}
\end{threeparttable}
%\end{minipage}
\end{table*}
The values show that the two different methods give very similar results (within a
factor typically of 2). In particular, the virial masses are slightly larger
than the masses computed using the conversion factor.
The two methods used here to determine the masses of the clouds are in
reasonable agreement and no clear evidence for a systematic difference between
the two determinations can be claimed at this point. There has been much
discussion of how well CO intensity can be converted to total molecular gas
mass in external galaxies using Galactic conversion factors. However, in the
present study we are resolving individual molecular clouds rather than
averaging over the entire distribution of clouds. The fact that the masses
derived by the two different methods agree within %\cut{the errors}  
a factor of 2, in spite of the different assumptions made for the two 
determinations,
suggests that there may not be a significant problem in this case. 

We also computed the surface density, $\Sigma$=M/($\pi$R$^2$), of each
identified cloud, using both the virial masses and the masses derived with the
conversion factor. The values are given in Table \ref{tab_param} (Col. 11 and
12).

\subsection{Continuum emission from the molecular clouds}\label{mma}

It may appear surprising that we do not detect 
1.3mm continuum emission from the
clumps seen in the CO emission or from the compact HII
regions seen with the VLA (see Fig.~\ref{canali} and \ref{co_cont}). 
As far as the HII regions are
concerned, we note that extrapolation to 230 GHz of the Johnson \& Kobulnicky
(2003) flux of the strongest centimeter continuum source, assuming optically
thin free-free emission, yields expected flux densities of 2 mJy or roughly 2
$\sigma $ at our sensitivity.  Thus it is not surprising that our observations
do not detect the free-free emission from these source, especially if
partially resolved.

%The most plausible interpretation is that we are observing dust emission from
%a compact source but in this case, it is surprising that the observed emission
%does not coincide with any of our CO clumps.  If interpreted as dust emission,
%we can infer the total gas mass $M_{g}$ from the observed flux density using
To compute the expected millimeter continuum flux from the dust, associated 
with the CO clumps (with gas mass $M_{g}$), we can use the equation:
\begin{equation}
 M_{\rm g}=\frac{ S_{\nu}\, d^2}{\kappa(\nu)\, B_{\nu}( T_{\rm d})} 
\label{gas-dust}
\end{equation}
(\cite{henning97}), assuming optically thin thermal dust emission and
isothermal conditions. $B_{\nu}$($T_d$) is the Planck function at the assumed
dust temperature T$_d$, $d$ the distance of the object and $\kappa(\nu)$ is the
dust opacity per gram of gas. We adopt opacity equal to 0.005 cm$^2$~g$^{-1}$
(Andr\'{e} et al. 2000).
For the derived masses of the individual CO clumps in Table~\ref{tab_param},
we estimate typical 1.3~mm continuum fluxes of 1-2 mJy (or 1-2 $\sigma$), 
for a range of temperatures $T_{\rm d}$ between 20~K and 50~K.
Therefore, it is not surprising that also from the CO clumps we do not detect 1.3~mm continuum emission.
%the continuum flux at
%1.3~mm converts into a gas mass between $5 \, 10^6 \, M_{\odot}$ and $1.6 \,
%10^7 \, M_{\odot}$.

%This mass is larger by a factor of 2-5 than the masses derived for individual
%CO clumps in Table~\ref{tab_param}, for these we estimate typical 1.3~mm
%continuum fluxes of 1-2 mJy according to Eq.~\ref{gas-dust} (or 1-2 $\sigma $).
Incidentally, we note that the continuum emission from dust may be a very
good tracer of dense and massive molecular clouds.
Indeed, an unresolved
clump of mass $M_{\rm cl}$ may be more easily detected in  continuum emission than
in CO.  This is because, for a source smaller than the synthesised beam, one
has for the ratio of continuum flux $S_{\rm d}$ to CO flux $S_{\rm CO}$ that:
\begin{equation}
\frac{S_{\rm d}}{S_{\rm CO}}\, = \, \frac{\kappa(\nu)\,M_{\rm cl}}{\pi\,\Delta \nu \,R_{\rm cl}^2}
\end{equation}
where $R_{\rm cl}$ is the clump radius and $\Delta \nu$ is the CO line width.
The latter in a virialised situation is proportional to $(M_{\rm cl}/R_{\rm
cl})^{0.5}$ and so one concludes that the ratio of continuum to CO fluxes
varies as $R_{\rm cl}^{-1.5}$ for an unresolved clump. We assume in the above
that the CO emission is optically thick and thermalised whereas the dust
emission is optically thin. For a cloud of radius 10 pc, the ratio is almost
ten times larger than the ratio for structures of the size of the beam ($R_{\rm
cl} \sim 40$~pc). This ratio can thus be significant if one is, for example,
observing something similar to the "Massive Molecular Aggregates" postulated by
Kobulnicky \& Johnson (2000), which are expected to have masses of order 10$^7
M_{\odot}$ and dimensions of order a few parsecs. Such objects would have 
millimeter fluxes of a few mJy and are barely out of reach for our
observations. Clearly, verifying this
speculation requires higher frequency, sensitivity and angular resolution
submillimeter observations of the continuum from Henize~2-10.

%________________________________________________________________

\section{Discussion}\label{discussion}

\subsection{Dynamical properties of the molecular gas}\label{dinamica}

An important aim of our study was to examine the kinematics of the molecular
clouds for comparison both with the velocity field of the stars (see e.g.
Marquart et al.  2007) and with earlier work on the gas kinematics by
Kobulnicky et al. (1995). This is useful also for comparison with the molecular
gas mass estimates derived in the previous section.

Thus in order to investigate the rotation curve in the inner region of
Henize~2-10 near the site of massive star formation and compare it with the
results of Kobulnicky et al. (1995), we extracted a position-velocity diagram
by taking a 4$^{\prime\prime}$ wide slice of our SMA CO($J$=2-1) data cube,
passing through the brightest optical starburst region, at the same position
angle as Kobulnicky et al. (1995) (PA=130$^{\circ}$), which they find to be the
direction of the steepest velocity gradient.  The orientation of the
position-velocity cut is shown in Fig.\ref{co_cont}.
\begin{figure}
\centering
\includegraphics[scale=0.4, angle=-90]{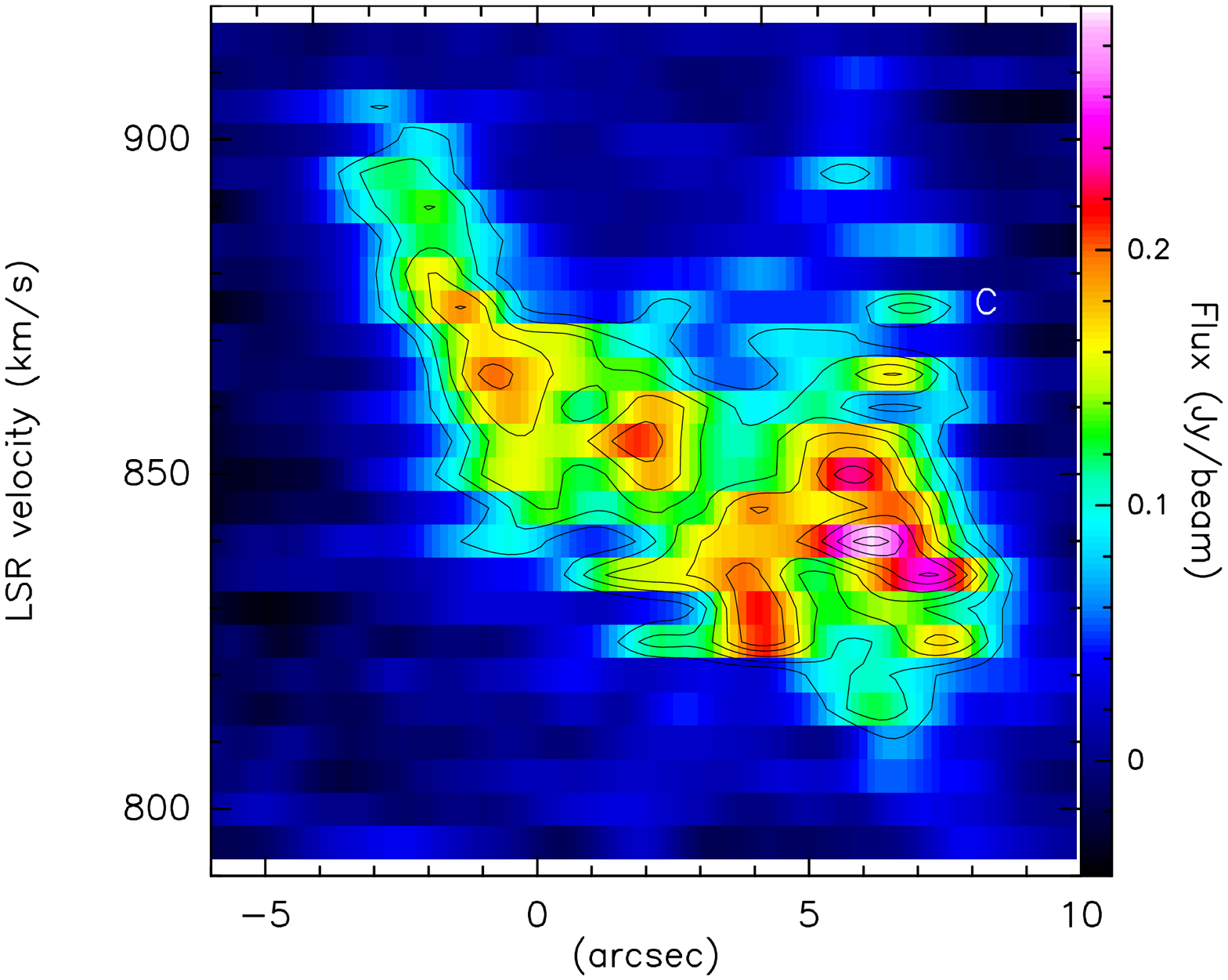}
\caption{SMA CO($J$=2-1) position-velocity diagram of Henize~2-10 obtained from
a 4$^{\prime\prime}$ wide slice through the data cube, at a position angle of
130$^{\circ}$ (see Fig.~\ref{co_cont}), passing through the center of the map:
08$^{\rm h}$36$^{\rm m}$15$.\!\!^{\rm s}$20, 
-26$^\circ$24$^\prime$34$.\!\!^{\prime\prime}$0 (J2000). 
The horizontal axis represents the positional
offset in arcsec with respect to this central position. Contour levels are
from 25\% to 95\% in step of 10\% of the peak.}
\label{posvel}
\end{figure}
The position-velocity diagram, shown in Fig. \ref{posvel}, reveals two
structures of which the innermost shows spatial velocity variations of 60
km~s$^{-1}$ on a scale of 7 arcseconds.  This velocity gradient implies a
dynamical mass of $1.6 \, 10^7/(\sin{i})^2$ $M_{\odot}$, within a radius of
150~pc, where $i$ is the inclination of the object  with respect to the line of
sight ($i$=90$^{\circ}$ means an edge-on rotating structure).

Kobulnicky et al. (1995) detected in their OVRO CO($J$=1-0) data 
a dynamically distinct feature, which was denoted by the letter ``C''.
This CO feature, around 875 km~s$^{-1}$ (see Fig.~\ref{posvel}), is also
detected in the SMA CO($J$=2-1) integrated map between 875 and 890 km~s$^{-1}$,
at the position $\alpha_{2000} = 8^h36^m15.\!\!^{s}5$ and $\delta_{2000} =
-26^{\circ}24^{\prime}39^{\prime\prime}$.
%\cut{The CO feature around 875 km~s$^{-1}$, denoted by the letter ``C'', is also
%detected in the SMA CO($J$=2-1) integrated map between 875 and 890 km~s$^{-1}$,
%at the position $\alpha_{2000} = 8^h36^m15.\!\!^{s}5$ and $\delta_{2000} =
%-26^{\circ}24^{\prime}39^{\prime\prime}$.  This dynamically distinct feature is
%seen in the OVRO CO($J$=1-0) data by Kobulnicky et al. (1995).}
%\cut{The} 
Its SMA CO deconvolved size is about $1.\!\!^{\prime\prime}8 \times 2.\!\!^{\prime\prime}3$,
which corresponds to a linear size of $80\times 100$~pc. Its total integrated
flux density is 4 Jy~km~s$^{-1}$, consistent with the measurement by
Kobulnicky et al. (1995), implying a molecular mass of $1.5 \, 10^6 \,
M_{\odot}$, using the same method described in section \ref{parameters} (see
Eq.~\ref{mmol}). Assuming a deconvolved velocity width of 14.1~km~s$^{-1}$ and a radius of
1$^{\prime\prime}$, we estimate for this feature a virial mass of $1.8 \,
10^6 \, M_{\odot}$ (see Eq.~\ref{virial}).
This cloud is not listed in Table~\ref{tab_param} since
%\cut{in the data cube, it is below the
%detection threshold that we have set for Clumpfind.}
its peak flux density is below the detection threshold (15~$\sigma$, see \S\ref{clumpfind})
that we have set for the output given by Clumpfind.

It is interesting to compare the derived dynamical mass with the estimated
stellar mass within the newly formed super star clusters and the mass of the
molecular gas in the region. From our SMA CO($J$=2-1), the mass of the
molecular gas within this region is around $10^7$ $M_{\odot}$, which means
that if the inclination of the molecular rotation axis is 90$^{\circ}$
(edge-on), the newly formed stars must contain at least 40\% of the total mass
in the inner 150~pc radius. If the inclination is lower, the total dynamical
mass is larger and, given that the total gas%\cut{s} 
mass is determined by our SMA
observations, the fraction of the total dynamical mass comprised by the stars
will be larger.

Conti \& Vacca (1994) present {\it Hubble Space Telescope} images which show 9
UV-bright knots in the inner 90~pc. %\cut{The} 
They estimate a total mass of newly formed
stars of $2.7\, 10^6 \, M_{\odot}$. 
The upper limit to the stellar mass from our SMA observations is $6\, 10^6\, M_{\odot}$
(see above), derived for a region which is about 3 times larger in radius than the one
considered by Conti \& Vacca (1994). This means that, extrapolating the estimated stellar mass from 
Conti \& Vacca (1994) to the dimensions of 
our region, one obtains a stellar mass value between $2.7\, 10^6 \, M_{\odot}$
and $2.4\, 10^7 \, M_{\odot}$, respectively assuming that the stars are concentrated in the inner 90~pc, 
observed by Conti \& Vacca (1994), or uniformly distributed over a region 
which is 3 times larger in radius. 
Therefore, the picture presented in Conti \& Vacca (1994) is consistent with the estimates which we discuss above.
%\cut{The region considered by us is about 3 times larger 
%but, very roughly, this seems consistent with the dynamical mass which we discuss above.}
%This result is a lower limit to the mass of
%the stars for a comparison with our region, \new{which is almost three times larger in size (150~pc radius)}, 
%but it seems to be in agreement with the scenario we derive\new{d above} from %\cut{our SMA CO data}
%\new{the kinematics of the CO clouds}.

Another way to estimate the total mass involved in the whole region and how it
is distributed between the molecular gas and the stars is to compute the
(statistical) velocity dispersion, $\sigma_{\rm V}$, of the clouds identified
in the SMA CO($J$=2-1) emission and derive the total mass of the region,
assuming virial equilibrium.  From Eq.~\ref{virial}, taking the FWHP
velocity ($\sqrt{8\,ln(2)} \times \sigma_{\rm V}$), we derived a total mass of
$1.2 \, 10^8$ $M_{\odot}$.  Given that the total gas mass derived from our SMA
CO($J$=2-1) data is about 4-6~$ 10^7$ $M_{\odot}$ (derived from the molecular 
and virial masses respectively; see Table \ref{tab_param}), we estimate again that the
stars should account for $\sim$30-50\% of the total mass, without taking into account 
the contribution from atomic gas, which is difficult to estimate at these angular scales.

Our result is also in agreement with the OVRO CO($J$=1-0) motions measured by
Kobulnicky et al. (1995).  Their study reveals a rotation of the CO gas on
large scales, with a velocity gradient which is about half of the gradient
derived from our SMA CO data. They find that the low resolution of their
observations ($6.\!\!^{\prime\prime}5 \times 5.\!\!^{\prime\prime}5$ beam size)
gives only an ``average'' rotation curve smoothed over the beam and decreases
the observed velocity gradient, which is consistent with what we observe.
Their estimated dynamical mass is $3.2 \, 10^6/(\sin{i})^2$ $M_{\odot}$, which
implies that the stellar component may contain up to half of the dynamical mass
in the inner region, in agreement with our result.

The gas velocity field is also consistent with Marquart et al. (2007). They
find spatial velocity variations up to 60 km s$^{-1}$ for the ionized
interstellar medium, but no systematic trend in the velocity of the stars and
conclude that the stellar kinematics is governed by random motions.

%__________________________________________________________________

\subsection{The physical properties of Clouds associated with SSCs}\label{physicalprop}

Some views of cluster formation expect prestellar cores to have the same
physical properties as the young stellar clusters they will eventually produce. 
%Figure \ref{tan} %(Tan 2007) 
%shows the surface density against mass, for star clusters and interstellar clouds. 
%\begin{figure}
%\centering
%\includegraphics[scale=0.45]{Tan.ps}            %Tan_ultimo.ps    Tan.ps
%\caption{Surface density, $\Sigma = M/(\pi R^2$), versus mass, $M$, 
%for star clusters and interstellar clouds. The large yellow circle represents the position 
%for integrated molecular mass and average densities of compact starburst galaxies hosting 
%SSCs (data from literature). The cyan circle is the position of the SMA CO($J$=2-1) clouds
%in Henize~2-10.}
%\label{tan}
%\end{figure}
%Indeed, as shown in the picture, 
Indeed, in the Galactic context the Infrared Dark Clouds (IRDCs), considered to
be possible high-mass pre-stellar cores and revealed by absorption of the
Galactic diffuse infrared background (Perault et al. 1996; Egan et al. 1998),
have similar surface densities as clusters associated with young massive stars
in our Galaxy (Tan~2008).  Therefore, in the extragalactic context, the
expectation is to find extremely dense molecular clouds associated with SSCs.
However, the limits imposed by the resolution and sensitivity of the existing
instruments make difficult the task of testing this scenario and resolving the parent dense
cores of the individual SSCs.
A 5~$\sigma$ detection of an unresolved clump of 10$^6$~$M_{\odot}$ requires 
a sensitivity of $\sim$0.1~mJy~beam$^{-1}$, assuming a temperature of 20~K, still a 
challenge for the current millimeter interferometric facilities.

The clouds, resolved in the CO(2-1) emission with the SMA, have masses and  
surface density values (see Table~\ref{tab_param}) of the same order of magnitude of the galactic 
GMCs ($M\sim 10^4-10^6\,M_{\odot}$, $\Sigma \sim 0.035$~g~cm$^{-2}$; 
Solomon et al. 1987), but systematically higher.
For a virialised cloud, surface density is a measure of 
pressure and it is comprehensible that the GMCs in Henize~2-10 are at
higher pressure than in the Galaxy. Moreover due to our limited
angular resolution, our column density estimates may well be considerable
underestimates. It seems reasonable therefore that, as in the Milky Way, clusters
in Henize~2-10 form from small parsec sized clumps at much
higher pressure than the rest of the GMC. %\cut{These clumps however
%likely contain a larger fraction of the GMC mass than in the Galaxy.}
Using SMA, we have made a first step towards
resolving the molecular clouds associated with the SSCs and towards testing for the presence
of very dense and compact molecular clouds, predecessors of the SSCs. 

\begin{figure}
\centering
\includegraphics[scale=0.45, angle = -90]{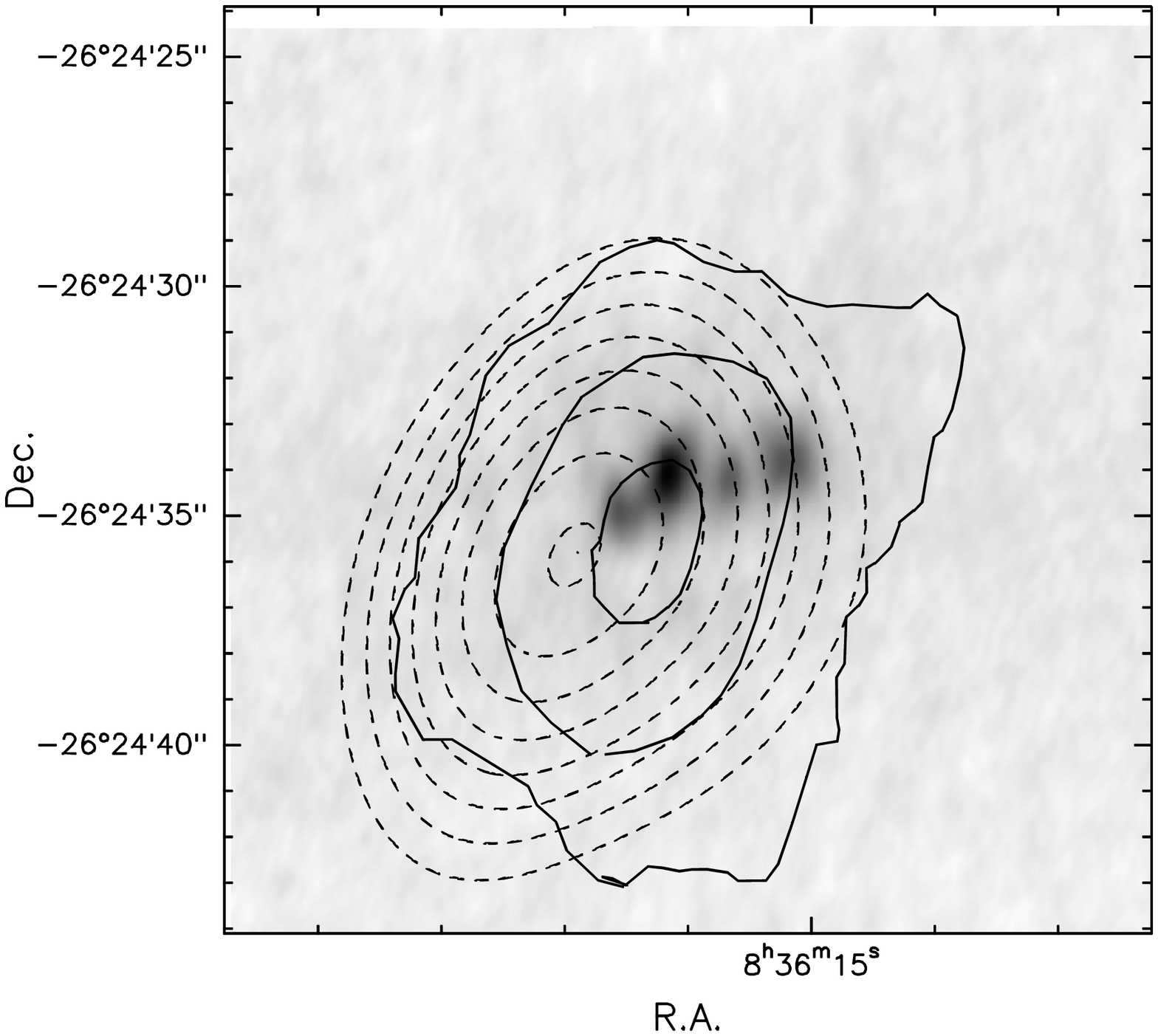}
\caption{HCO$^{+}$($J$=1-0) integrated map from Imanishi et al. (2007) (beam
size of $5.\!\!^{\prime\prime}5 \times 10.\!\!^{\prime\prime}8$, solid
contours) and SMA CO($J$=2-1) emission (dashed contours) smoothed to the
resolution of the HCO$^{+}$($J$=1-0), overlaid on the VLA 3.6~cm continuum in
grey scale (Johnson \& Kobulnicky 2003).}
\label{imani}
\end{figure}

The density traced by the CO($J$=2-1) line is limited (critical density
$n_{cr}\sim2.7 \, 10^3$~cm$^{-3}$) and to probe high density molecular gas
($n_{\rm H_2}>10^4$~cm$^{-3}$), which is expected to be associated with star
(SSCs) forming regions, higher density tracers like HCO$^{+}$ and HCN are
needed. Indeed, Henize~2-10 is known to show strong HCO$^+$ and HCN emission
(Imanishi et al. 2007; our own IRAM 30~m spectrum presented in this paper),
which reflect the spatial distribution of the dense molecular gas. The
interferometer observations of Imanishi et al. (2007) have a beam size of
$5.\!\!^{\prime\prime}5 \times 10.\!\!^{\prime\prime}8$. To compare our SMA
CO($J$=2-1) with the NMA maps of Imanishi et al. (2007), we have smoothed our maps to
the same angular resolution. The comparison is shown in Fig.~\ref{imani}.
Given that the position calibration uncertainties for the interferometric data are 
much less than $1^{\prime\prime}$, 
the overlay of the CO($J$=2-1) and HCO$^+$($J$=1-0) emissions seems to reveal a shift
between the two molecular emissions, which is almost $2^{\prime\prime}$.
This is surprising on one side because
one would expect the HCO$^+$($J$=1-0), being a tracer of high density gas, to
be associated to the peaks of the CO($J$=2-1) molecular emission. On the other
hand, the CO distribution is complex in both position and velocity and it is
likely that only some of the molecular clouds may be sufficiently dense to
show strong HCO$^{+}$($J$=1-0) emission and this might produce the shift
between the centers of the two integrated emissions.  Besides, there might be
also an effect due to the different opacity of the two lines.  Finally, the
HCO$^+$($J$=1-0) peak from Imanishi et al. (2007) appears to be closer to the
SSCs in Henize~2-10 than our SMA CO($J$=2-1) peak. This could mean that the
high density gas is mainly close to the SSCs.  
%Nevertheless, we also warn that
%this comparison may be altered by the limited (u,v) coverage of our SMA
%observations, which begin to resolve out a significant fraction of the emission
%on scales comparable to the Imanishi et al. (2007) resolution (see \S\ref{mmcontinuum}).

Indeed, the sensitivity and resolution of the HCO$^{+}$($J$=1-0) observations
of Imanishi et al. (2007) are not adequate in space and velocity to resolve the
dense gas peaks associated with the SSCs.  It is thus necessary to observe high
density tracers at higher resolution to resolve the molecular clouds out of
which the SSCs formed.

%_______________________________________________________________

\section{Summary and Conclusions}\label{summary}

In this paper we have presented new single dish and interferometric 
observations of the molecular gas and millimeter continuum of the 
starburst region in the dwarf galaxy Henize~2-10. 

Our observations confirm earlier detections of the $^{13}$CO and HCN(1-0)
line emission. The relatively strong detections of high density molecular
tracers associated with the young Super Star Clusters confirm that this
galaxy is undergoing vigorous star formation and is an ideal laboratory 
to study extragalactic starbursts. 

In this context, our CO(2-1) and 1.3mm continuum SMA observations, with a
linear resolution of $60-80$~pc, represent a first attempt to resolve the
parent molecular clouds out of which SSCs may form. We reveal a rich 
population of molecular clouds with estimated masses and densities in the
upper range of those measured in our Galaxy. The molecular gas accounts for 
approximately half of the total mass in the inner region of the galaxy, while
the young stellar clusters account for the remaining mass.

We find possible evidence that the super star clusters are associated
with very massive and dense molecular cores, but our observations and
tracers do not allow us to confirm unambiguously their presence. New
higher angular resolution and sensitivity observations of high density
tracers and (sub-)millimeter continuum  are 
required to clarify this possibility.

%We have presented new, high angular resolution CO($J$=2-1) SMA observations of
%the blue compact dwarf galaxy Henize~2-10.  The aim of our observations was to test
%the expectation that the formation of SSCs requires very dense molecular
%cores. The SMA observations allow us to resolve the molecular emission in 14
%structures, with masses between 2 and $7\, 10^6 \, M_{\odot}$, which may be
%GMCs.  We detect a 5 mJy 1.3 mm continuum source which is not coincident with
%either CO clouds or the VLA continuum emission. Assuming this to be due to
%dust emission, we infer a mass between $5.3 \, 10^6 \, M_{\odot}$ and $1.6 \,
%10^7 \, M_{\odot}$, corresponding to a range of temperature $T=20-50 K$.  We
%see rotation structures in our SMA CO map and infer a dynamical mass of $1.6\,
%10^7/(\sin{i})^2 M_{\odot}$ within a radius of 150 parsec.  Even if our data
%do not yet allow us to probe the range of densities expected for the proto-SSC
%clouds, our initial findings suggest that we may be seeing forerunners of
%super starclusters (SSCs).  Future high angular resolution observations of
%higher density tracers, HCO$^+$ and HCN, may allow to provide a final answer. 

%%______________________________________________________________
%
%%\section{Conclusions}
%
%   \begin{enumerate}
%      \item 
%      \item 
%      \item 
%   \end{enumerate}

\begin{acknowledgements}
All the observations presented in this paper were performed by the
observatories staff in service mode. We thank all the astronomers and
operators on duty for delivering excellent observations and for delivering
calibrated data packages very rapidly. In particular we thank Thomas Stanke
and Carlos de Brueck for the Science Verification observations with APEX,
Pierre Cox and Stephane Leon for the IRAM-30m DDT observations and
Ray Blundell and Quizhou Zhang for scheduling additional SMA observations
in the compact array. We also thank Kelsey Johnson and Jonathan Tan for
providing the VLA 3.6cm map and for insightful discussions.
\end{acknowledgements}

\end{document}